\documentclass[prd,twocolumn,footinbib]{revtex4}
\usepackage{draftcopy}
\usepackage{amsfonts}
\usepackage{amssymb}
\usepackage{graphicx}
\usepackage{pstricks}

\newcommand{\be}{\begin{equation}}
\newcommand{\bea}{\begin{eqnarray}}
\newcommand{\ee}{\end{equation}}
\newcommand{\eea}{\end{eqnarray}}
\newcommand{\bpi}{\begin{picture}}
\newcommand{\bce}{\begin{center}}
\newcommand{\epi}{\end{picture}}
\newcommand{\ece}{\end{center}}

\begin{document}

\title{Decoherent Neutrino Mixing, 
Dark Energy and Matter-Antimatter Asymmetry}

\author{Gabriela Barenboim$^a$}
\author{Nick E. Mavromatos$^{b}$}
\affiliation{$^a$Departamento de F\'\i sica Te\'orica and IFIC, Centro Mixto, 
Universidad de Valencia-CSIC,
E-46100, Burjassot, Valencia, Spain. \\
$^b$King's College London, University of London, Department of Physics, 
Strand WC2R 2LS, London, U.K.}

\begin{abstract}
A CPT violating decoherence scenario can easily account for all
the experimental evidence in the neutrino sector including LSND.
In this work it is argued that this framework can also 
accommodate the Dark Energy content
of the Universe, as well as the observed matter-antimatter asymmetry.
\end{abstract}



\maketitle

In a previous work\cite{bm}, henceforth referred to as I, 
we have discussed a phenomenological way 
of accounting for the LSND results~\cite{lsnd}
on evidence for antineutrino oscillations ${\overline \nu}_e \to 
{\overline \nu}_\mu$, but with lack of 
corresponding evidence in the neutrino sector,
by means of invoking CPT Violating (CPTV) decoherence, due to 
quantum gravity. Indeed, quantum decoherence in matter propagation
occurs when the matter subsystem interacts with an `environment'~\cite{ehns},
according to the rules of open-system quantum mechanics. 
At a fundamental level,
such a decoherence may be the result of 
propagation of matter in quantum gravity space-time backgrounds
with `fuzzy' properties, which may be responsible for violation
of CPT symmetry\cite{cpt} in a way not necessarily related to mass differences
between particles and antiparticles.
As demonstrated in I, 
it is possible to fit all the available neutrino data,
including the results from LSND and Karmen\cite{karmen} experiments,
not by enlarging the neutrino sector or implementing 
CPTV mass spectra for neutrinos, but by invoking 
a CPTV difference in the decoherence parameters between particle and 
antiparticle sectors in three generation neutrino models (we refer the reader to the
original work for technical details). 
{}From this point of view, then, 
the LSND result would evidence CPT violation in the sense of  
different decohering interactions between particle and antiparticle
sectors, while the mass differences (and widths) 
between the two sectors remain the same. From I
it became clear that both mixing and decoherence, the latter
in the antineutrino sector only, were necessary to account for all 
the available 
experimental information,
including LSND and Karmen results\cite{lsnd,karmen}. 
Mixing, in the sense of non trivial
mass differences between energy eigenstates, 
was important, since pure decoherence,
that is absence of any mass terms in the Hamiltonian, 
was not sufficient to fit the data.
However, this does not mean that the mass terms are necessarily
of conventional origin. As stressed in I, the Hamiltonian
appearing in the decoherent evolution 
should be viewed as an ``effective'' one, receiving 
possible contributions from the environment as well. 
In this sense, one cannot exclude the 
possibility that some contribution to the neutrino masses 
have a quantum-decoherence origin, as a result of interactions 
with the foam, as happens for example when neutrinos interact with matter
and the mass differences get modified (and mass degeneracies lifted) as
a result of the interaction. 
Such an effect will disentangle neutrino masses from standard 
electroweak symmetry breaking scenaria.
As we shall discuss below, this is an important feature that will allow
us to associate the recently claimed amount of dark energy 
in the Universe by means of astrophysical 
observations\cite{snIa,wmap} to our
decoherent neutrino mass differences.

The (energy depending) decoherence parameters needed to account for 
all the experimental information, 
$\gamma_{1}  \sim 10^{-18} \cdot E,~\gamma_{3} 
\sim 10^{-24}/E $, found in 
our sample point in I, call for an explanation within a 
consistent theoretical framework. 
To this end, the reader should first observe that, for energies of a few GeV,
which are typical of
the pertinent experiments, 
such values are not far from  
$\gamma_j \sim \Delta m_{ij}^2$. 
If our conclusions
survive the next round of experiments, and therefore if 
MiniBOONE experiment~\cite{miniboone} confirms previous LSND claims,
then this may be a significant result. One would be tempted~\cite{bm} to 
conclude that if the above estimate holds,  
this would probably mean that 
the neutrino mass differences
might 
be due to quantum gravity decoherence. Theoretically it is still 
unknown how  neutrinos acquire a mass, or what kind of mass 
(Majorana
or Dirac) they possess. 
Thus, if the above model turns out to be right we might then have, 
for the first time in low energy physics, 
an indication of a direct detection of a quantum gravity effect, which 
disguised itself as an induced decohering neutrino mass difference. 
Notice that in our  model only antineutrinos have non-trivial 
decoherence parameters,
while the corresponding quantities in the neutrino sector vanish.  
This implies that there may be a single cause for mass differences,
the decoherence in antineutrino sector, compatible 
with common mass differences in both  sectors. 

In what follows  we  will make this assumption, namely that decoherence 
effects, due to interactions with the foam, contribute 
to the Hamiltonian terms in the evolution of the neutrino density 
matrix, and result in  
neutrino mass differences in much the same way as the celebrated
MSW effect\cite{msw}, responsible for a neutrino mass 
splitting due 
to interactions  with a medium. 
Indeed, when 
neutrinos travel through matter, the 
neutral current contribution to this interaction, proportional 
to -$G_F n_n/\sqrt{2}$, with $G_F$ Fermi's weak interaction constant, 
and $n_n$ the neutron density in the medium, 
is present for {\it both} $\nu_e$ and 
$\nu_\mu$ (in a two flavour scenario), while the charged current contribution, given by 
$\sqrt{2}G_Fn_e$, with $n_e$ the medium's electronic density, 
is present only for $\nu_e$. The flavour eigenstates 
$\nu_{e,\mu}$ can then be expressed in terms of fields
${\tilde \nu}_{1,2}$ with definite masses ${\tilde m}_{1,2}$ 
respectively, with a mixing angle ${\tilde \theta}$, the tilde notation
indicating the effects of matter. The tilded quantities are diagonalised
with respect to the Hamiltonian of $\nu_e$,$\nu_\mu$ in the presence 
of non-trivial matter media, and one can find the 
following relations between vacuum (untilded) and medium parameters\cite{msw}
${\rm sin}^22{\tilde \theta} \simeq {\rm sin}^22\theta 
\left(\frac{\Delta m^2}{\Delta {\tilde m}^2}\right)$, with  
$\Delta {\tilde m}^2 = \sqrt{(D -\Delta m^2{\rm cos}2\theta )^2 +
(\Delta m^2{\rm sin}2\theta)^2}, D=2\sqrt{2}G_Fn_e k$. From this 
we observe that the medium-induced effects in the 
mass splittings are proportional 
to the electronic density of the medium and in fact, even if the neutrinos
would have been mass degenerate in vacuum, such a degeneracy would 
be lifted by 
a medium.

To get a qualitative idea of what might happen with 
the foam, one imagines a similar mixing for neutrinos, as a result 
of their interaction with a quantum-gravity decohering foam situation.
As a result, there are {\it gravitationally-induced effective masses} 
for neutrinos,
due to flavour dependent interactions of the foam, which are in principle
allowed in quantum gravity. In analogy (but we stress that 
this is only an analogy) 
with the MSW effect, the gravitationally-induced 
mass-splitting effects are expected now to be proportional
to $G_N n_{\rm bh} k$,
where $G_N =1/M_P^2$ is Newton's constant,
$M_P \sim 10^{19}$ GeV is the quantum gravity scale, 
and $n_{bh}$ is a ``foam'' density of appropriate 
space time defects (such as Planck
size black holes {\it etc.}), whose interaction with the neutrinos
discriminates between flavours, in an analogous way 
to the matter effect. Neutrinos, being 
electrically neutral can indeed interact 
non-trivially with a space time foam, and change flavour as a result
of such interactions, since such processes are allowed by quantum gravity. 
On the other hand, due to electric
charge conservation of microscopic black holes, 
quarks and charged leptons, 
cannot interact non-trivially with the foam. 
In this spirit, one can imagine a microscopic 
charged black-hole/anti-black-hole pair being 
created by the foam vacuum.
Evaporation of these black holes 
(probably at a slower rate than their neutral counterparts, due to
their near extremality~\cite{ebh})
can produce 
preferentially $e^+e^-$ pairs (lighter than muons), 
of which the positrons, say, are absorbed into the microscopic
event horizons of the evaporating 
charged anti-black hole.
This leaves us with a stochastically fluctuating (about a mean value)
electron (or more general charge) density, $n^c_{\rm bh}(r)$, 
induced by the gravitational foam, 
$\langle n^c_{\rm bh} (r)\rangle = n_0 \ne 0$, 
$\langle n^c_{\rm bh} (r)n^c_{\rm bh} (r')\rangle \ne 0$,
which, in analogy
with the electrons of the MSW effect in a stochastically
fluctuating medium\cite{loreti}, can interact non-trivially only
with $\nu_e$ but not with the $\nu_\mu$, in contrast to 
neutral
black holes which can interact with all types of neutrinos\cite{details}. 
We assume, of course, that the 
contributions 
to the  
vacuum energy that 
may result from such 
emission and absorption processes
by the black holes in the foamy vacuum  
are well within the known limits. For instance,
one may envisage supersymmetric/superstring models of space-time foam, 
where such contributions may be vanishingly small\cite{west}.
The mean value (macroscopic) part, $n_0$, of 
$n^c_{\rm bh}(r)$, assumed time independent, will contribute to the 
Hamiltonian part of the evolution of the neutrino density matrix, $\rho$.   
In analogy with the (stochastic) MSW effect\cite{msw,loreti},
this part yields space-time foam-induced 
mass-squared splittings for neutrinos:
\begin{equation}
\langle\Delta m^2_{\rm foam}\rangle 
\propto G_N \langle n^c_{\rm bh}(r)\rangle k
\label{mswgrav}
\end{equation}
with non trivial quantum fluctuations ($k$ is the neutrino momentum scale).  
To ensure a constant neutrino mass one may consider the case where 
$\langle n^c_{\rm bh}(r)\rangle$, which expresses the average number 
of virtual particles emitted from the foam with which the neutrino
interacts, is inversely proportional to 
the (neutrino) momentum. This is reasonable, since the faster the neutrino, 
the less the available time to interact with the foam, and hence 
the smaller the number of foam particles 
it interacts with. Such flavour-violating foam 
effects would also contribute to decoherence
through the quantum fluctuations of the foam-medium 
density\cite{loreti,details},
by means of induced non-Hamiltonian 
terms in the density-matrix evolution. 
Such effects assume a double commutator structure\cite{loreti,details,adler}
and are due to {\it both}, the fluctuating parts of the foam density, as well
as the effects of the mixing (\ref{mswgrav}) 
on the vaccum energy. 
Indeed, as we shall show below, 
neutrino flavour mixing leads
to a non-trivial contribution to the vacuum energy, in a non-perturbative 
way suggested in \cite{vitiello2}.
Hence, 
such effects are necessarily CPT violating\cite{mlambda}, 
in the sense of entailing an evolution 
of an initially pure neutrino quantum state to a 
mixed one due to the presence of the Hubble horizon
associated with the non zero cosmological constant,
which prevents pure asymptotic states from being well defined. 
In that case, CPT is violated  
in its strong form, that is CPT is not a well-defined operator, 
according to the theorem of \cite{wald}.

For convenience we shall discuss explicitly the two-generation
case. The arguments can be extended to three generations, 
at the expense of an increase in mathematical complexity,
but will not affect qualitatively the conclusions
drawn from the two-generation case. 
The arguments are based on the observation\cite{vitiello1} 
that in quantum field theory, which by definition requires an 
infinite volume limit, in contrast to 
quantum mechanical treatment
of fixed volume\cite{pontecor}, the neutrino {\it flavour} states are 
{\it orthogonal} to the 
{\it energy} eigenstates, and moreover they define 
two inequivalent vacua
related to each other by a {\it non unitary} transformation $G^{-1}(\theta,t)$:
$|0(t)\rangle_f = G^{-1}_\theta (t)|0(t)\rangle_m$,
where $\theta$ is the mixing angle, $t$ is the time, and the suffix f(m)
denotes flavour(energy) eigenstates respectively, and 
$G^{-1}_\theta (t) \ne G^{\dagger}_\theta (t)$ is a non-unitary operator
expressed in terms of energy-eigenstate neutrino free fields $\nu_{1,2}$\cite{vitiello2}: $G_\theta (t) = {\rm exp}\left(\theta \int d^3x [\nu_1^\dagger (x)
\nu_2 (x) - \nu_2^\dagger (x)
\nu_1(x)]\right)$.
A rigorous mathematical
analysis of this problem has also appeared in \cite{hannabus}. 
As a result of the non unitarity of $G^{-1}_\theta (t)$, there is a 
Bogolubov transformation\cite{vitiello1} 
connecting the creation and annihilation operator 
coefficients appearing in the expansion of the appropriate  
neutrino fields of the energy or flavour eigenstates. 
Of the two Bogolubov coefficients appearing in the
treatment, we shall concentrate on $V_{\vec k} = 
|V_{\vec k}|e^{i(\omega_{k,1} +\omega_{k,2})t} $, with 
$\omega_{k,i}=\sqrt{k^2 + m_i^2}$,  
the (positive) energy of the neutrino energy eigenstate $i=1,2$ 
with mass $m_i$. This function is related to the condensate 
content of the flavour vacuum, in the sense 
of appearing in the expression of an appropriate 
non-zero number operator of the flavour vacuum\cite{vitiello2,hannabus}:
$_f\langle 0 |\alpha_{{\vec k}, i}^{r \dagger}\alpha_{{\vec k}, i}^{r}
|0\rangle_f = _f\langle 0 |\beta_{{\vec k}, i}^{r \dagger}
\beta_{{\vec k}, i}^{r}|0\rangle_f = {\rm sin}^2\theta |V_{\vec k}|^2$ 
in the two-generation scenario~\cite{vitiello1}. 
$|V_{\vec k}|$ has the property of vanishing for $m_1=m_2$, it has a maximum 
at the momentum scale $k^2 =m_1m_2$, and for $k \gg \sqrt{m_1m_2}$ 
it goes to zero as:
\begin{equation}
|V_{\vec k}|^2 \sim \frac{(m_1 - m_2)^2}{4|{\vec k}|^2}, \quad 
k \equiv |{\vec k}| \gg \sqrt{m_1m_2}
\label{bogolubov}
\end{equation}
The analysis of \cite{vitiello2} argued that the flavour vacuum $|0\rangle$, 
is the
correct one to be used in the calculation of the average vacuum energy, 
since otherwise the probability is not conserved\cite{henning}.
The energy-momentum tensor of a Dirac fermion field in the Robertson-Walker
space-time background can be calculated straightforwardly in this formalism. 
The flavour-vacuum average value of its temporal $T_{00}$ component,
which yields the required contribution to the vacuum energy 
due to neutrino mixing, is\cite{vitiello2}:
\begin{eqnarray}
&& _f\langle 0|T_{00} |0\rangle_f 
=\langle \rho_{\rm vac}^{\rm \nu-mix}\rangle \eta_{00} \nonumber \\
&& = \sum_{i,r}\int d^3 k \omega_{k,i}\left(_f\langle 0 |\alpha_{{\vec k}, i}^{r \dagger}\alpha_{{\vec k}, i}^{r}|0\rangle_f + 
_f\langle 0 |\beta_{{\vec k}, i}^{r \dagger}\beta_{{\vec k}, i}^{r}|0\rangle_f\right) 
= \nonumber \\
&& 8{\rm sin}^2\theta \int_0^K d^3k (\omega_{k,1} +  
\omega_{k,2})|V_{\vec k}|^2.
\label{vacener}
\end{eqnarray}
where $\eta_{00}=1$ in a Robertson-Walker (cosmological) metric background.
The momentum integral in (\ref{vacener}) is cut-off 
from above at a certain scale, $K$ 
relevant to the physics of neutrino mixing. In conventional approaches,
where the mass generation of neutrino occurs at the electroweak phase 
transition, this cutoff scale can be put on the electroweak
scale $K \sim 100 $ GeV, but this yields unacceptably large contributions to
the vacuum energy. An alternative scale has been suggested 
in \cite{vitiello2}, namely $K \sim \sqrt{m_1m_2}$ as the
characteristic scale for the mixing. 
In this way these authors 
obtained a phenomenologically acceptable value
for $\langle \rho_{\rm vac}^{\rm \nu-mix} \rangle $. 

In our case we 
shall use a different cutoff scale, 
which allows for some analytic estimates of (\ref{vacener})
to be derived, as being  
mathematically consistent with the 
asymptotic form of (\ref{bogolubov}), which  
is valid in a regime of momenta $k \gg \sqrt{m_1m_2}$.
This cutoff scale is simply given by the sum of 
the two neutrino masses, 
$K \equiv k_0 = m_1 + m_2$, 
is compatible with our decoherence-induced mass difference scenario,
and also allows for a mathematically consistent analytic estimate of the 
neutrino-mixing contribution to the vacuum energy in this framework.
For hierarchical neutrino models, for which $m_1 \gg m_2$, we have 
that $k_0 \gg \sqrt{m_1m_2}$, and thus, if we assume that the 
modes near the cutoff contribute most to the vacuum energy 
(\ref{vacener}), which is clearly supported by the otherwise divergent 
nature of the momentum integration, and take into account the 
asymptotic properties of the function $V_{\vec k}$, which are safely valid
in this case, we obtain: 
\begin{eqnarray} 
&&\langle \rho_{\rm vac}^{\rm \nu-mix}\rangle \sim 
8\pi{\rm sin}^2\theta (m_1 - m_2)^2 (m_1 + m_2)^2\times \nonumber \\ 
&&\left(\sqrt{2} + 1 +{\cal O}(\frac{m_2^2}{m_1^2})\right) \propto {\rm sin^2}\theta (\Delta m^2)^2
\label{darkneutr}
\end{eqnarray}
in the limit $m_2 \ll m_1$. 
For the (1,2) sector, the corresponding $\Delta m^2 $ is given by the solar neutrino
data and is estimated to be $\Delta m^2_{12} \simeq 10^{-5}$ eV$^2$, resulting in a
contribution of the right order.
This dependence of the cosmological constant 
on the square of the neutrino mass-squared difference
has been conjectured in I, and was ``derived'' here following the
flavour/mixing quantum field theoretic treatment of \cite{vitiello2}.
In this way 
the cosmological constant $\Lambda$ is  
elegantly expressed in terms 
of the smallest (infrared, $\Delta m^2$) and the largest 
(ultraviolet, $M_P^2$) Lorentz-invariant mass scales available.
The choice of the cutoff $k_0 \sim m_1 + m_2$ is 
consistent with our conjecture on the decoherence origin of the neutrino
mass difference, due to interaction with the foam medium (\ref{mswgrav}).
Indeed, for momenta $k \sim m_1 + m_2$,
which have been argued above to be the dominant contribution to the 
dark energy component (\ref{darkneutr}), the induced mass splittings 
become $\Delta m^2_{\rm foam}  
\sim G_N \langle n^c_{\rm bh}\rangle (m_1 + m_2) $ from which 
$m_1 - m_2 \sim \langle n^c_{\rm bh}\rangle /M_P^2 $. If we assume there are 
${\cal N}_c $ charged foam-induced 
objects per Planck volume, $V_P \sim M_P^{-3}$ 
then,  ${\cal N}_{c, \rm max} \sim m_1 - m_2 /M_P$. 
For realistic neutrino mass values 
this is very small, 
indicating that in such scenaria, a tiny amount of black holes
in the foamy vacuum suffices to produce observable effects in neutrino 
physics. It goes without saying of course that none of the above statements
should be considered as a rigorous derivation. Nevertheless, we think that
the above arguments are non trivial and we believe they may be related 
with an actual theory of  quantum gravity. 
Notice that the above way of deriving the neutrino-mixing contribution 
to the dark energy is independent of the usual perturbative loop arguments,
and, in this sense, the result (\ref{darkneutr}) 
should be considered as  exact 
(non perturbative), if true.

Some important remarks are now in order. 
First of all, our choice of cutoff scale was such that the 
resulting contribution to the cosmological constant depends on the 
neutrino mass-squared differences and not on the absolute mass,
and hence it is independent of any zero-point energy, in agreement
with energy-driven decoherence models~\cite{adler}.
For us, it is curved space physics that is  responsible for 
lifting the mass degeneracy of neutrino mass eigenstates and create the
``flavour'' problem. 
This is an important point, which may serve as motivation (not proof)
behind such a cutoff ``choice'', which we conjecture 
is a physical ``necessity''.
We have argued above that such a cutoff ``choice'' 
is a natural one from the point of 
view of quantum-gravity decoherence-induced mass differences. 
Detailed models of this fall way beyond the purposes of this brief note.
Nevertheless, we believe 
that the above-demonstrated self-consistency of this cutoff choice
within the remit  
of our toy model of space time foam 
is intellectually challenging and encouraging for further studies
of this important issue. 

It should be noted at this stage that our
considerations above are based on the suggestion (which is not beyond doubt) 
of ref. \cite{vitiello1} on a Fock-like quantisation of 
the flavour space. There is still controversy in the literature 
regarding the physical meaning of such quantum flavour states~\cite{giunti},
in particular it has been argued that, although such states are mathematically
elegant and correct constructions, 
nevertheless they lead to no observable consequences. 
However, 
in view of the results of \cite{vitiello2} and of the present work,
such an argument may not be correct, since the 
mass-squared difference contribution to the cosmological constant
is an observable (global) consequence of the Fock-like flavour space 
quantisation.
The presence of a time independent cosmological constant (\ref{darkneutr})
in the flavour vacuum, which notably is not present if one uses instead the
mass eigenstate vacuum, 
implies an asymptotic future event horizon for the emerging 
de-Sitter Universe. 
The flat-space time arguments of 
\cite{giunti} for the flavour space field theory 
cannot then be applied, at least naively, and the problem 
of quantisation
of the Fock-like flavour space is equivalent to the (still elusive) 
quantisation of field theories in (curved) de-Sitter space times.

In such a case one cannot define properly 
asymptotic states, and hence a scattering matrix. 
This will lead to decoherence, in the sense of a modified temporal 
evolution for matter states. 
For instance, string theory considerations\cite{mlambda} 
suggest that the temporal evolution of the matter
density matrix $\rho$ in such a de-Sitter Universe, will be decoherent:
$\partial_t \rho = i[\rho, H] + :\Lambda g_{\mu\nu}[g^{\mu\nu}, \rho]:$,
where $\Lambda $ is the cosmological constant, given in our case 
by (\ref{darkneutr}), and  $: \dots :$ denotes quantum ordering.
Notice that the decoherent non-Hamiltonian term is proportional
to (a quantum version of) the conformal anomaly (trace of the stress tensor)
of the de-Sitter universe.
For antisymmetric
ordering, one obtains a 
double commutator structure $[g_{\mu\nu}, [g^{\mu\nu}, \rho]]$,
which when considered between, say, energy eigenstates yields 
variances of the metric field $(\Delta g_{\mu\nu})^2$, expressing
quantum fluctuations of the space time geometry, as a result of the
interactions of neutrinos with the foam.
The terms proportional to $\Lambda$  
lead in general to a decoherent 
evolution  of a pure quantum mechanical state to 
a mixed one. According to the general arguments of \cite{wald}, then,
one should expect in this case a strong form of CPT Violation, in the sense
that the CPT operator is not well defined. 
This may lead to different decoherent parameters eventually between
particles and antiparticles, reflecting the different ways 
of interaction with the foam between the two sectors. In other words,
it is possible that the variance $({\overline \Delta g_{\mu\nu}})^2$
in the antiparticle sector is much larger than the corresponding
one in the particle sector. See however \cite{details} for the 
suppression of this second effect in our case, where the 
foam-density-fluctuation terms may be held 
responsible for the leading contributions to 
decoherence. Nevertheless, due to the presence of the $\Lambda$ term
there will be a mixed state CPTV description.

We now notice that, in the case of 
(anti)neutrinos passing through stochastic media, including 
space time foam,  
there are additional contributions to decoherence, which may offer
a natural explanation of the decoherence parameters of I.
An important source of decoherence in such media is due to the 
uncertainties in the energy $E$ and/or the oscillation 
length $L$ of the (anti)neutrino beam. 
In fact, it can be shown~\cite{ohlsson}
that if one averages the standard oscillation 
probabilities $P_{\nu_\alpha \to \nu_\beta}$
over Gaussian distributions for $E$ and/or $L$ with a variance $\sigma^2$,
the result is equivalent to neutrino decoherence models, in the sense
of the time dependent profile of the associated 
probability being identical to that of a completely-positive  
decoherence model.
One finds for $n$ flavours~\cite{ohlsson}, 
\begin{eqnarray} 
&& \langle P_{\alpha\beta} \rangle = \delta_{\alpha\beta} 
- 
2 \sum_{a=1}^n\sum_{\beta=1, a<b}^n{\rm Re}\left(U_{\alpha a}^*
U_{\beta a}U_{\alpha b}U_{\beta b}^*\right)\times \nonumber \\
&& \left( 1 - {\rm cos}(2\ell \Delta m_{ab}^2)
e^{-2\sigma^2(\Delta m_{ab}^2)^2}\right) -
\nonumber \\
&& 2 \sum_{a=1}^n \sum_{b=1, a<b}^n {\rm Im}\left(U_{\alpha a}^*
U_{\beta a}U_{\alpha b}U_{\beta b}^*\right) \times \nonumber \\
&& {\rm sin}(2\ell \Delta m_{ab}^2)
e^{-2\sigma^2(\Delta m_{ab}^2)^2}
\label{avprobs}
\end{eqnarray}
where $U$ is the mixing matrix 
$\ell \equiv \langle x \rangle$, 
$\sigma = \sqrt{\langle (x - \langle x \rangle)^2} \equiv (L/4E)r$, 
and $x = L/4E$.
The form is identical to that of decoherence, as becomes evident
by noting that the 
exponential damping factors can be written 
in the form $e^{-\gamma_j L}$
with $t=L~(c=1)$, and 
decoherence parameters $\gamma_j$ of order: 
$2\sigma^2_j (\Delta m^2)^2 = \gamma_j L$, from which  
$\gamma_j = \frac{(\Delta m^2)^2}{8E^2}Lr^2_j$.
There are various 
scenaria that restrict the order of $\sigma$.
In general, the
acceptable bounds on $\sigma$ may be divided in two 
major categories, depending on the form of the uncertainties~\cite{ohlsson}:
$\sigma_j \simeq \Delta x \simeq \Delta_j\frac{L}{4E}
\le \frac{\langle L \rangle}{4 \langle E \rangle}
\left(\frac{\Delta_j L}{\langle L \rangle} 
+ \frac{\Delta_j E}{\langle E \rangle}\right)$, or 
$\sigma_j 
\le \frac{\langle L \rangle}{4 \langle E \rangle}
\left([\frac{\Delta_j L}{\langle L \rangle}]^2 + 
[\frac{\Delta_j E}{\langle E \rangle}]^2\right)^{1/2}$. 
In three generation models the values of the length and energy uncertainties
may vary between flavours, and also between neutrinos and antineutrinos,
as a result of the intrinsic CPT violation, hence the subscript $j$ 
in the above formulae (for 
antiparticle sectors it is understood that $j \to {\overline j}$).
From the above considerations it becomes clear that, for 
$L \sim 2E/\Delta m^2$, which is characteristic for oscillations, 
one has decoherence parameters $\gamma_j \sim (\Delta m^2/E)r_j^2$.
It is interesting to estimate first the order of 
decoherence induced by conventional physics, 
for instance decoherence induced by uncertainties
in the measured energy of the beam due to experimental limitations.
For long base line, atmospheric or cosmic neutrino experiments,  
where $\Delta L/L$ is negligible, and $\Delta E/E \sim 1$ 
such decoherence 
parameters are found 
at most of order $\gamma \sim 10^{-24}$ GeV, 
for the relevant range of energies, 
and they diminish with 
energy, vanishing formally when $E \to \infty$, which seems to be a 
general feature of conventional matter-induced  decoherence 
effects~\cite{ohlsson}. 
 
To obtain the decoherence parameters of the best-fit model of 
I it suffices to choose for the antineutrino sector 
$r_{\overline 3}=r_{\overline 8} \sim \Delta E /E \sim 1 $,
and $r_{\overline 1}^2=r_{\overline 2}^2 \sim 10^{-18} \cdot E^2/\Delta m^2 $.
As seen above, the decoherence 
parameters exhibiting a $1/E$ energy dependence
could be attributed 
to conventional energy uncertainties 
occuring in the beam of the (anti)neutrinos. 
However, the parameters proportional to $E$, if true, may be attributed
to exotic physics.

The fact that $r_j$ in general receives contributions from 
both length and energy uncertainties provides a natural
explanation for the different energy dependence of the 
decoherence parameter of the model of I in the antiparticle
sector. Indeed, having identified $r_{\overline 3}=r_{\overline 8}$ 
as decoherence 
induced by `conventional-looking'
energy uncertainties in the antineutrino sector, 
it is natural to assume that the $\gamma_3=\gamma_8 \propto E$ decoherence
is due to genuine quantum gravity effects, increasing with energy, 
which are associated with metric tensor quantum fluctuations. 
This is achieved provided we assume that 
$r_{\overline 3}^2=r_{\overline 8}^2 \sim (\Delta L/L)^2$,
i.e. these decoherence coefficients are 
predominantly oscillation-length-uncertainty driven,  
and take into 
account that variations in the invariant length may be caused by metric
fluctuations, since $L^2 =g_{\mu\nu}L^\mu L^\nu$, implying
$(\Delta g_{\mu\nu})^2 \sim (\Delta L)^2/L^2$, in order of magnitude.
To obtain the best fit results of I, then, 
for $L \sim 2E/\Delta m^2$, one needs quantum-gravity induced
metric fluctuations in the antineutrino sector 
of order $ (\overline \Delta g_{\mu\nu})^2 \sim 10^{-18} L \cdot E$. 
The increase with energy
is not unreasonable, given that the higher the energy of the antineutrino
the stronger the back reaction onto space time, and hence the 
stronger the quantum-gravity induced metric fluctuations. The factor
$10^{-18}$ may be thought of as being of order $E/M_P$, with
$M_P \sim 10^{19}$ GeV the Planck mass, although 
alternative interpretations may be valid (see 
discussion on possible cosmological interpretations at the end of the 
article). The increase with $L$ is not uncommon in stochastic models
of quantum foam, where the decoherence `medium' 
effects build up with the distance the (anti)particle
travels~\cite{west}.
We also mention at this stage that, 
apart from 
these effects, 
in stochastic models
of foam there are additional contributions to decoherence, arising
from the fluctuations of the density of the medium.
These too can mimic the effects of the best-fit model of I
in the antineutrino sector, as discussed in some detail in \cite{details},
but their $L$-dependence is different from that of the above effects.
Comparison between short and long baseline experiments, therefore,
may differentiate between the various decoherent contributions.  

Unfortunately at present, we lack a detailed microscopic model
of space-time 
foam, and hence the above  
considerations should be treated with caution. 
Nevertheless, 
we think that the above plausibility arguments, as well as those 
in \cite{details}, attempting to 
explain the order and the energy dependence of the 
decoherence parameters of I are not unreasonable.

In view of our conjecture on 
the quantum-gravity origin of the mass differences 
between neutrino flavours, supported by the above analysis, 
we should stress 
that we are clearly dealing here with an {\it interacting} theory
on (highly) curved space times, and the ordinary procedure of 
quantisation
needs to be completely rethought. 
If the mass difference is time independent,
then, as argued above, one cannot follow standard methods of 
free-field quantisation~\cite{vitiello2,giunti}, due to
the non-well defined nature of the scattering matrix, leading to CPT 
Violation. Based on this, one might even conjecture a situation in
which the foam flavour vacuum is relaxing to equilibrium as the 
cosmic time elapses, in such a way that the asymptotic value of the 
neutrino mass-squared difference vanishes, 
and a proper set of asymptotic states
can be defined. These are very interesting, and highly non-trivial issues,
that we would like to bring to the attention of the reader
at this stage, merely to argue that, 
if our conjecture on the quantum-gravity origin  of the 
neutrino mass difference is valid, 
then 
the flavour mixing issue is far 
from being resolved, and certainly it cannot be treated 
in the way addressed in 
\cite{giunti}. 
We hope to be able to address such questions in a 
more detailed manner in a future publication.  

As a final remark we would like to draw a connection
between our decoherence scenario and the matter-antimatter
asymmetry of the Universe. As is well known,
sphaleron transitions occurring at and after the electroweak (EW) phase
transition induce violations of $B+L$\cite{Kuzmin:1985mm}, which efficiently wipe out
any pre-existing $B+L$ asymmetry. Leptogenesis models evade
this problem by generating an early asymmetry in $L$, which
is then converted to a baryon asymmetry by the $B-L$ conserving
sphaleron processes \cite{Riotto:1999yt}. 
To avoid sphaleron dilution of $B+L$, and
to satisfy the Sakharov conditions \cite{Sakharov:dj}
for baryogenesis,
standard leptogenesis  models require strongly out-of-equilibrium 
processes and
new sources of $CP$ violation beyond the Standard Model.
Our model of decoherence on the contrary provides
a novel and extremely economical mechanism to generate
the observed baryon asymmetry, through a process of
equilibrium electroweak leptogenesis (the fact that it violates
CPT obviates the
need for two of the three Sakharov conditions, namely the requirements
of out-of-equilibrium and CP violating processes).
Put it more formally, by breaking CPT
and thus the axioms of quantum field theory, we have violations
of the index theorem that relates the Chern-Simons winding number
of the sphaleron configuration to a change in $B+L$. 
It is difficult to do a precise calculation of this effect,
but it is easy to derive an order of magnitude estimate.
In \cite{lopez} the asymmetries between
semileptonic decays of $K_0$ and those of ${\overline K_0}$ 
turned out to depend linearly on dimensionless decoherence 
parameters such as  ${\widehat \gamma} = \gamma/\Delta \Gamma$; 
in the parametrization of Ellis et al. in \cite{ehns}, 
where $\Delta \Gamma=\Gamma_L - \Gamma_S$ was a characteristic
energy scale associated with energy eigenstates of the kaon system.
In fact, the dependence was such that the decoherence 
corrections to the asymmetry were of order ${\widehat \gamma}$ 
in complete positivity scenaria, where only one decoherence parameter,
$\gamma > 0$ was non zero.
In similar spirit, in our case of lepton-antilepton number asymmetries,
one expects the corresponding asymmetry to depend, to leading order, 
linearly on the quantity 
${\widehat { \gamma}} = \gamma/\sqrt{\Delta 
m^2}$, since the quantity $\sqrt{\Delta m^2}$ is the characteristic 
energy scale in the neutrino case, playing a role analogous to 
$\Delta \Gamma$  in the kaon case. The only difference from the kaon 
case,
is that here, in contrast to the kaon asymmetry results, there are no
zeroth order terms, and thus the result of the matter-antimatter asymmetry
is proportional to the dimensionless decoherence parameter 
$\widehat{\gamma}$, which we are going to take as the larger of the two
dechorence parameters of our model in \cite{bm}, {\it i.e.}  $
\widehat{\gamma} \to \widehat{\gamma_1} = 10^{-18} \cdot E/\sqrt{\Delta m^2}$.

Cutting this long story short, the matter-antimatter asymmetry in the 
Universe is estimated to be 
${\cal A} =  \frac{\langle \nu \rangle - \langle {\overline \nu} 
\rangle}{\langle \nu \rangle + \langle {\overline \nu} \rangle}
\simeq  \, \widehat{\gamma_1} \, \simeq  10^{-6}$~\footnote{The numerical
coefficient $10^{-18}$ on $\gamma$ may be thought of as the 
ratio $T/M_P$ with $T$ the temperature, whose value gets frozen at 
the EW symmetry breaking temperature.}. 
Thus, $B= \frac{n_\nu -\bar{n_\nu}}{s} \sim 
\frac{{\cal A} n_\nu }{g_* n_\gamma}$ with $n_\nu $ ($\bar{n_\nu} $) 
the number density of (anti) neutrinos, $n_\gamma$ the number density of 
photons and $g_*$ the effective
number of degrees of freedom  (at the temperature where the
asymmetry is developed) which depends on the exact matter content
of the model but it ranges from $10^2$ to $10^3$ in our
case. This implies a residual baryon asymmetry of order $10^{-10}$,
roughly the desired magnitude.

In this work we have used a minimal model of CPT violating decoherence,
able to explain all observations in neutrino experiments, 
in an attempt to account
for the vacuum energy of the Universe as well as for its matter-antimatter
asymmetry. This extremely simplified model 
which incorporates just two (decoherence) parameters
to  the standard three  generation scenario  
is able through some educated guesses
to get numbers in the right ball-park 
for these two apparently unrelated quantities.
Obviously we are in need of a detailed theoretical model of foam
before definite conclusions are reached in these important issues,
but we think that our discussion in this work places
neutrino physics in a quite novel perspective
that is worth of further study.

\section*{Acknowledgements}

We thank J. Lykken and G. Gounaris for discussions.
The work of of GB is supported by CICYT, Spain,
under grant FPA2002-00612, while that 
of NEM is partly supported by the European Union 
(contract HPRN-CT-2000-00152). 
NEM thanks the Department of 
Theoretical Physics of the University of Valencia for hospitality
during the final stages of this work.

\end{document}